# Keyboard for inputting Chinese language
## -A study based on US patents

**Umakant Mishra**

Bangalore, India

umakant@trizsite.tk

http://umakant.trizsite.tk

**Contents**

# 1. Introduction

## 1.1 Technique of inputting Chinese character

As the structure of Chinese characters are very different from the relatively simple alphabetic system of western languages, it is very difficult to input Chinese characters into computer quickly and conveniently. There are a few existing systems which include those based on the "PinYin" (phonetic) system, a combination of the PinYin system and character form techniques, whole character encoding, stroke input encoding, and stoke form encoding.

### The PinYin (Phonetic) system

The PinYin encoding system presents several advantages over other systems. First, it is already part of the curriculum of China's primary schools and many people are therefore familiar with the system. Besides, PinYin uses English characters as its encoding characters. Therefore it is a simple matter to directly encode Chinese characters by using standard English character computer keyboards.

However, since the output of the PinYin encoding system consists of Chinese characters, and not the PinYin forms of those characters, there are too many degenerate cases. Generally some software processing is used to compensate this effect. For a language such as Chinese, with its many dialects and slangs, there are simply too many Chinese characters for the user to remember or pronounce correctly.

### The whole character system

The whole character keyboard is based on the use of a special hardware keyboard and not the standard English language keyboard. The input speed of this keyboard is comparatively low and is not suitable for wide use.

### Stroke input encoding system

The stroke input encoding system is easy to learn. But it presents too many degenerate cases and is not suitable for entering large amounts of data.

### Stoke from encoding system

The stoke form encoding system is based on a process of first determining the radicals of the characters, then encoding them according to an established set of rules. The stoke form systems developed to date, however, have suffered from the lack of a suitably developed theoretical structure. Therefore, the system still suffers from a large number of synonym codes and slow input speed. It is also difficult to use.

### 1.2 Problems relating to inputting Chinese characters

- The Chinese character strokes and symbols are so different and so complicated that they can be stored and grouped in a wide variety of ways.

- The conventional keyboard does not support the pictorial characters in Chinese language. There are 3000 to 6000 commonly used pictorial Chinese characters (Hanzi).

- There are too many dialects and slangs in Chinese which makes the language more difficult to remember and pronounce correctly.

- In "PinYin" type of system, the user needs to know two aspects of Chinese characters, viz., their pronunciation and their written form. This is inconvenient, as the user has to do two operations.

- There are too many generates of PinYin system of encoding. Besides, in Taiwan, Chinese characters are denoted with different phonetic symbols, not English letters. Therefore, PinYin faces a special barrier in Taiwan and other Asian countries.

- Unlike English, there are no boundaries between words in Chinese text. For example, a sentence can be a contiguous string of ideograms, where one or more ideograms may form a word, without spaces between "words".

- Some encoding systems, while high in speed, are poorly described, thereby leading to confusion.

- Some encoding systems fail to use to the standard 26-character English character keyboard. For example, the "whole characher system" uses a special hardware keyboard.

## 2. Inventions of inputting Chinese characters

### 2.1 Method and apparatus for inputting radical-encoded Chinese characters (Patent 5119296)

**Background problem**

There remains a problem of inputting Chinese characters into computers because of the complicated structure of Chinese characters. The different systems of inputting Chinese characters such as PinYin system, stroke input encoding system, stoke from encoding system etc. have their own limitations.

**Solution provided by the patent**

Zheng et al. provided a system (patent 5119296, issued June 1992) of computer encoding system for Chinese characters. The system analyzes different Chinese characters and identifies six basic strokes. Besides 26 most frequently used radicals are selected and classified into these six categories. Each of these radicals is depicted by English character key-positions on a standard computer keyboard.

| THE STROKE FORM | — | \| | / | \ | ㄱ | ㄴ |
|---|---|---|---|---|---|---|
| NAME OF THE STROKE | THE HORIZONTAL STROKE | THE VERTICAL STROKE | THE LEFT-SLANTING STROKE | THE RIGHT SLANTING STROKE | UP-RIGHT CURVING | LOWER-LEFT CURVING |
| THE STROKE NUMBER | 1 | 2 | 3 | 4 | 5 | 6 |
| VARIETIES OF THE STROKE | — . — | 1 . ↓ | / . ⁄ | ﹨ | — . — Z . ㄱ | ㄴ.l.﹨ L.ㄴ. |

The invention assigns English characters which denote the above single radicals as codes for Chinese characters such that each Chinese character is encoded by four or fewer English characters. The codes input through the keyboard are converted to Chinese by a converter.

**TRIZ based analysis**

The invention tries to use the regular English keyboard for inputting Chinese characters **(Principle-3: Local quality)**.

The Chinese characters in the regular script are classified into six basic strokes **(Principle-1: Segmentation)**.

The 26 most frequently used radicals are mapped on the 26 English character key positions on the keyboard **(Principle-28: Mechanics substitution)**.

**2.2 Keyboard for processing Chinese language text (Patent 5893133)**

**Background problem**

It is necessary to produce the characters of Chinese language with a conventional keyboard. However, the conventional keyboard does not support the pictorial characters as in Chinese language. There are 3000 to 6000 commonly used Chinese characters (Hanzi) and so many more rarely used. It is difficult to enter the Chinese characters through the keyboard for text processing.

**Solution provided by the invention**

Chenjun Julian Chen discloses a method of accurately and efficiently entering Chinese language through the keyboard (patent 5893133, assigned to IBM, April 99). The system has a keyboard with diacritic keys that permits the user to annotate each entered phonetic text syllable with a diacritic that indicates the tone of the syllable. The user enters phonetic Chinese (Pinyin and BPMF) into a computer system and the phonetic input is compared to a list of acceptable phonetic syllables and abbreviations. If the entered syllable is found on the list, the syllable is displayed on a phonetic portion of the graphical display. When the user presses a delimiting character the phonetic syllable is converted to proper hanzi character and displayed on the screen.

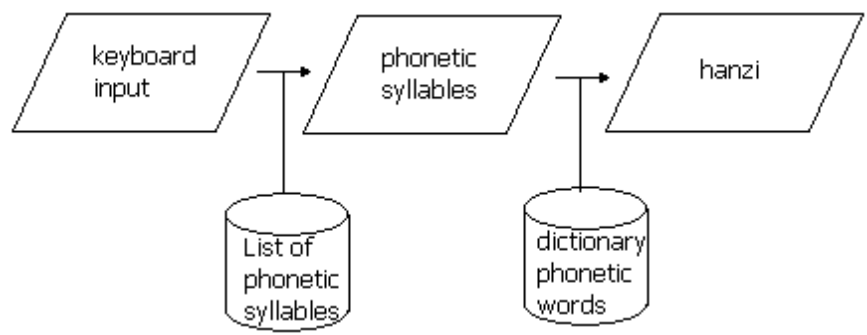

**TRIZ based analysis**

A standard keyboard should support easy input of all hanzi characters **(Ideal Final Result)**. But the number of hanzi are more than 10,000 which are not possible to generate out of such a small keyboard **(Contradiction)**.

The invention uses some special keys and key combinations to generate more characters out of the limited number of keys **(Principle-40: Composite)**

The invention uses a different intermediate form like phonetic Chinese which has less number of characters (about 1300) compared to hanzi **(Principle-24: Intermediary)**.

## 3. Summary and conclusion

The complexity of Chinese language and Chinese character have made people struggle to input Chinese characters into computer. Entering of Chinese characters and processing of Chinese language will remain a topic of continuous research and development for all times in future.